\documentclass[prb]{revtex4}
\usepackage{graphicx}
\usepackage{dcolumn}
\usepackage{amsmath}
\usepackage{latexsym}

\makeatletter
\def\btt#1{\texttt{\@backslashchar#1}}%
\DeclareRobustCommand\bblash{\btt{\@backslashchar}}%
\makeatother

\begin{document}

\preprint{HEP/123-qed}

\title[Short Title]{Effective Stochastic Resonance under Heterogeneous Amplitude of Noise \\}


\author{Shogo Torigoe$^1$}%

\author{Ryosuke Kawai$^1$}%

\author{Kazuhiro Yoshida$^2$}

\author{Akinori Awazu$^1$}%

\author{Hiraku Nishimori$^1$}%
\email{nishimor@hiroshima-u.ac.jp}
\affiliation{${}^1$Department of Mathematical and Life Sciences, Hiroshima University, Kagamiyama, Higashi-hiroshima 739-8526, Japan
}%
\affiliation{${}^2$Department of Mathematical Sciences,  Osaka Prefecture University, Sakai 599-8531,Japan
}%

\date{\today}

\begin{abstract}
Effective stochastic resonance (SR) is numerically and
 analytically studied using a model with coupled two particles exposed
 to heterogeneous, i.e., particles dependent, amplitude of noise.
Compared to previous SR models of single particle and to those of
 coupled two particles exposed to equivalent amplitude of noise, the
 present model exhibits a more intensive resonance of, at least, one
 particle exposed to the non-larger amplitude of noise with the assistance
 of another particle.
In a certain range of conditions, this effective resonance of one particle
 overwhelms the poor resonance of the other particle, meaning that
 heterogeneous amplitude of noise leads the system, not only locally but
 also in the average of the whole, to the effective SR.
\end{abstract}

\maketitle
\section{Introduction}
\bigskip
\noindent
As reviewed in \cite{14}, various kinds of phenomena have been recognized as 
stochastic resonance phenomena (SR) and have been studies with a variety
of methods\cite{0, 3, 19, 20, 15, 16, morishita, suzuki}.
Among theoretical models for SR, the simplest is that
described by the Langevin dynamics of a particle confined in a
double-well potential periodically deformed by an oscillating external
field,
\begin{eqnarray}{}
\frac{\partial x}{\partial t} &=& -\frac{\partial V(x)}{\partial x}+\xi(t), 
\end{eqnarray}
where,
\begin{equation}
V(x)=-ax^2+bx^4-A\cos(\Omega t)x \label{2},
\end{equation}
the last term of which indicates the oscillating external field.
Hereafter, we call the above model as the original SR model.

To quantify the efficiency of SR, the signal-to-noise ratio (SNR), the
Spectral Power Amplification (SPA) and other quantities have been
proposed.  
According to recent relating studies\cite{SPA}, here, we employ SPA as
the index of SR.  SPA is, roughly, the power ratio between the input and
the output signals, and is defined as, 
\begin{eqnarray}{}
\frac{\int^{\Omega+\delta}_{\Omega-\delta}
 S_{out}(\omega)d\omega}{\int^{\Omega+\delta}_{\Omega-\delta}
 S_{in}(\omega)d\omega},
\label{SPA}
\end{eqnarray}
where $\Omega$, $S_{in}(\omega)$ and $S_{out}(\omega)$ represent the
frequency of the input signal, power spectrums of input and output
signals, respectively.
The denominator means the power of the input signal (sinusoidal wave),
and the numerator is the corresponding power of the output signal.
Here, $\delta$ is taken to be larger than the peak width of the power spectrum. 
In our simulation, considering the resolution of the numerical Fourier
transformation, (\ref{SPA}) is approximated by  
\begin{eqnarray}{}
\frac{S_{out}(\Omega)}{S_{in}(\Omega)}.
\end{eqnarray}

A series of efforts to obtain the higher SPA or SNR have been conducted
through variations of the original SR model, one of which was made by
coupling two particles applied in a common double-well potential
periodically deformed by the external field as described by (\ref{2}).
Through the reasonable tuning of coupling strength, this coupled
particles model gives a higher SNR than the original SR
model\cite{1,referee1}. 
As another directional variation, cascade dynamics using one-way
coupling between oscillating sub-systems was simulated\cite{13}, that
exhibits a high synchronization to the oscillating external
field by adding a proper amplitude of noise even below Hopf bifurcation
threshold.
Note that most of the previous models of SR have treated systems in
which constituent elements are exposed to equivalent amplitude of noise,
whereas a few models have considered the heterogeneous, i.e., element
dependent, amplitudes of noise.
As discussed in one of such models\cite{morishita}, the
noise emerging in a living cell is not always equivalent through the
cell because the origin of noise in living systems is, in addition to
the thermal fluctuation, the local density fluctuation of
reacting molecules in each cell. 

Therefore, as one way for revising the preceding SR models considering
at the time the potential comparison to actual systems, we study SR
model of coupled two elements exposed to the heterogeneous amplitude of
noise, restricting the model as simple as
possible to focus only on the basic mechanism of the underlying system.

\section{Langevin Dynamics Model and Simulations}
The present model is described as, 
\begin{equation}
\begin{array}{l}
\displaystyle\frac{\partial x_1}{\partial t} = -\frac{\partial
 V(x_1)}{\partial x_1} +K(x_2-x_1)+\xi_1(t), \\[3mm]
\displaystyle\frac{\partial x_2}{\partial t} = -\frac{\partial
V(x_2)}{\partial x_2} +K(x_1-x_2)+\xi_2(t),
\label{1}
\end{array}
\end{equation}
where $V(x)$ is same as (\ref{2}), $a=8.0$, $b=0.25$, $A=10.0$ and
$K$ is coupling constant.
The quantities $\xi_1(t)$ and $\xi_2(t)$ are Gaussian white noise with
the relations,
\begin{equation}
\displaystyle \langle \xi_i (t) \rangle = 0,
 \hspace{1pc} \langle\xi_i(t )\xi_j(s)\rangle =
 2D_i\delta_{i,j}\delta(t-s) \hspace{2pc} (i,j \in \{ 1, 2 \}).
\end{equation}

It should be notified that noise amplitudes $D_1$ and $D_2$
are {\it independently} varied as a set of control parameters.
We numerically calculate (\ref{1}) and measure SPA for respective
particles with varying a set of parameters $(D_1, D_2, K)$.
The numerical integration is carried out using the 4th order Runge-Kutta
method for deterministic part and the Euler-Maruyama scheme for
stochastic part with time mesh $\Delta t = 0.005$.
All data are taken as the ensemble averages of $100$ simulations with
each run performed from $t=0$ to $t=102 \times 2 \pi/ \Omega$ under the
randomly chosen initial conditions of $x_i$ ($x_i \in \{ -\sqrt{a/2b},
\sqrt{a/2b}\}$), where we take away the dynamics from $t=0$ to $t=2
\times 2 \pi /\Omega$ to avoid the dependence to initial conditions.
In the following, we show the numerical outcomes. Specifically, SPA of
particle 1 named SPA1, and the average SPA of two particles named
ASPA=(SPA1+SPA2)/2, are focused on.

\begin{figure}[pbt]
\includegraphics[width=5.5in, bb=0 0 1285 774]{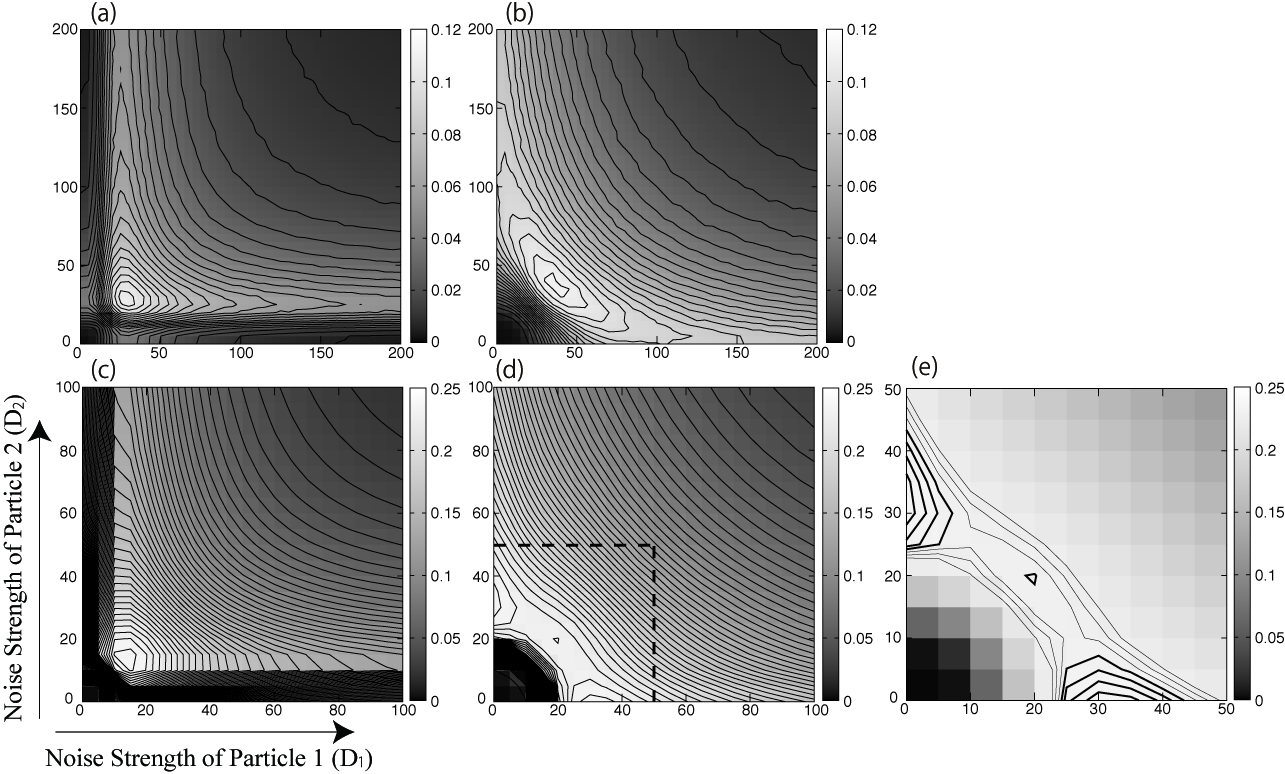}
\caption{Contour lines of ASPA in $D_1-D_2$ space for
 (a)$K=1.00,\:\Omega=\pi/4$, (b)$K=5.00,\:\Omega=\pi/4$,
 (c)$K=1.00,\:\Omega=\pi/64$, (d)$K=5.00,\:\Omega=\pi/64$, (e)extended figure of the left-lower part of (d).
The brighter tone means the higher value of ASPA. 
ASPA is symmetric with respect to the diagonal line $D_1=D_2$ from the
 definition of ASPA.
Only in the case of (d)( and (e)), the peaks of ASPA fall {\it off} the diagonal, equivalent amplitude of noise, line.
In (d),  contours of ASPA $\geq$ 0.230  are drawn by bolder lines and those of ASPA $ < $ 0.222  are abbreviated to focus on the maximum points
 }

\label{ASPA1}
\label{ASPA1-1}
\end{figure}
Fig.\ref{ASPA1} shows ASPA in $D_1-D_2$ parameter space
for different sets of ($K$, $\Omega$).
Note that Fig.\ref{ASPA1}(a)-(d) are symmetric with respect to the
diagonal line $D_1=D_2$ from the above definition of ASPA.
As the oscillation frequency of external field, two values,
  $\Omega=\pi/4$ and $\Omega=\pi/64$, are chosen, and as the coupling
  strength, $K=1.00$ and $K=5.00$ are chosen.
Fig.\ref{ASPA1}(a) and (b) show ASPA numerically obtained with $\Omega =
\pi/4$ in the cases of weak $(K=1.00)$ and strong $(K=5.00)$ couplings,
respectively. 
For both cases, the peaks of ASPA roughly falls on the line $D_1=D_2 >0$
 meaning that equivalent amplitude of noise leads the system to the most effective SR in $D_1-D_2$ space in these cases of $\Omega = \pi/4$.
With a slower oscillation $(\Omega=\pi/64)$ of external field, if
coupling strength is weaker than a critical value, the equivalent
amplitude of noise still gives the largest ASPA as seen in
Fig.\ref{ASPA1}(c).
On the other hand, as shown in Fig.\ref{ASPA1}(d) (and its enlarged view
Fig.\ref{ASPA1}(e)) for  a stronger
coupling ($K=5.00$) with $\Omega=\pi/64$, the peaks of ASPA in $D_1-D_2$
space are definitely located $\it off $ the equivalent amplitude line of
noise.  
This means that heterogeneous amplitude of noise leads the system to the
most effective SR in $D_1-D_2$ space in this case.

\begin{figure}[tb]
\includegraphics[width=5.5in,bb=0 0 779 289]{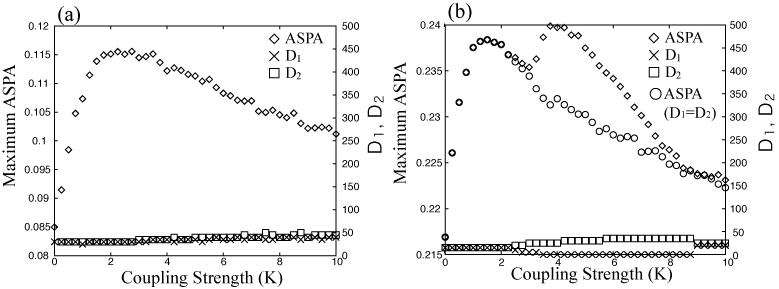}
\caption{Maximum ASPA obtained in each $D_1-D_2$ space with different
 $K$ value ($\diamond$), Maximum ASPA obtained under equivalent noise
 ($D_1 = D_2$) ($\circ$), and corresponding set $(D_1, D_2)$ ($\Box$ and
 $\times$) for (a) $\Omega=\pi/4$, (b) $\Omega=\pi/64$, here, only the sets
 satisfying $D_1 \leq D_2$ are drawn.}
\label{ASPA2}
\label{ASPA2-1}
\end{figure}

In Fig.\ref{ASPA2}, maximum ASPAs obtained at each $K$-fixed $D_1-D_2$
space (hereafter max-ASPA) are connected, varying $K$, as graphs.
The symbol $\diamond$ in Fig.\ref{ASPA2}(a) shows  the relation between  $K$ and
 max-ASPA in the case of $\Omega=\pi/4$. Here, SR of the system is
 reinforced as the coupling between particles is strengthened until
 max-ASPA reaches a peak value at finite $K$ ($K\simeq 2$).
In the same figure, the symbols, $\Box$ and $\times$ respectively give
the sets ($D_1$, $D_2$) realizing max-ASPA at each $K$.
As mentioned above, the value of ASPA is symmetric with respect to $D_1$
and $D_2$ from the definition of ASPA, therefore two sets of ($D_1$, $D_2$)
realizing max-ASPA are seen if $D_1 \neq D_2$.
Fig.\ref{ASPA2} (and Fig.\ref{SPA12}) show only the sets satisfying $D_1
\geq D_2$.
Two symbols $\Box$ and $\times$ roughly collapse on each
other in Fig.\ref{ASPA2}(a) indicating that max-ASPA is attained under
equivalent amplitude of noise in this case. 
Fig.\ref{ASPA2}(b) shows the case of slower oscillation $\Omega=\pi/64$.
Like in Fig.\ref{ASPA2}(a), the symbol $\diamond$ shows the relation
between $K$ and max-ASPA in each $K$-fixed $D_1-D_2$ space.
In this figure, max-ASPA graph consists of two parts divided
by a dip around $K \simeq 3$, each of which has a mound shape, and as
mentioned afterward, the right-side mound including the highest point in the
graph is mainly contributed by the particle exposed to the lower
amplitude of noise.
The symbols, $\Box$ and $\times$, in the same figure give the set
$(D_1, D_2)$ of noise amplitude with which max-ASPA is realized at
each $K$.
Obviously different from the case of $\Omega=\pi/4$ shown in
Fig.\ref{ASPA2}(a), $D_1$ and $D_2$ in Fig.\ref{ASPA2}(b) separate from
each other for $2.25 \leq K \leq 9.25$, within which interval the
highest value of the max-ASPA is realized.
Namely, the most effective SR  in $D_1-D_2$ space is realized when
heterogeneous amplitude of noise is applied to the system.	
To close up further the effect of heterogeneous amplitude of noise, we
plot in Fig.2(b) maximum ASPA obtained along $D_1=D_2$ line in each
K-fixed $D_1-D_2$ space. Compared with max-ASPA obtained without the
constraint  $D_1=D_2$ in the same figure, we see the right-side mound in
max-ASPA graph is definitely ascribed to the heterogeneity  of adding
noise.


\begin{figure}[tb]
\includegraphics[width=4.in, bb=0 0 871 775]{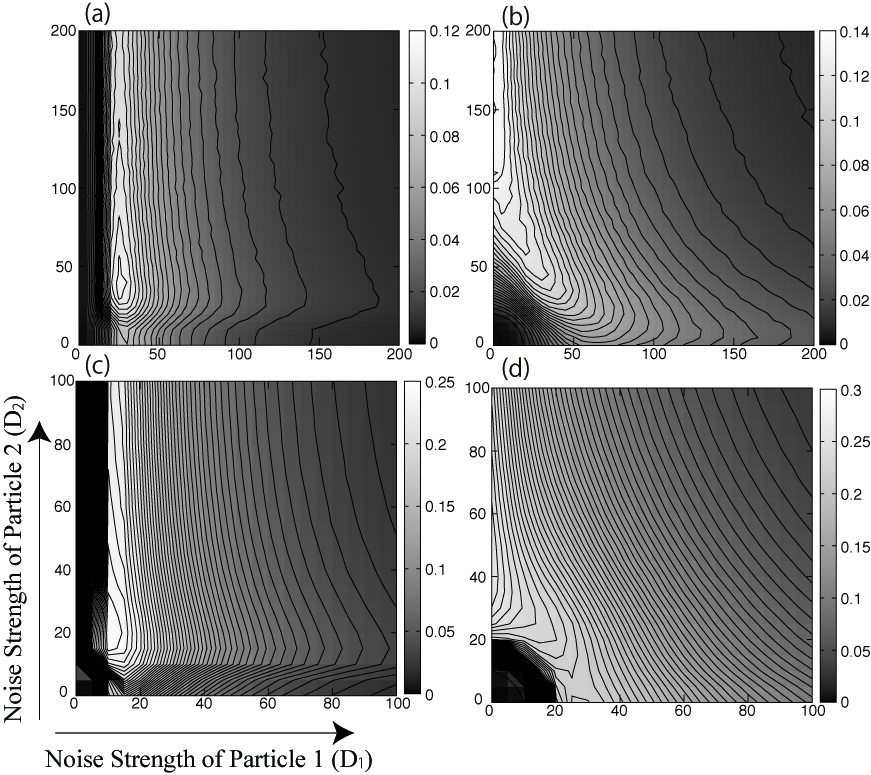}
\caption{Contour lines of SPA1 in $D_1-D_2$ space for
 (a)$K=1.00,\:\Omega=\pi/4$, (b)$K=5.00,\:\Omega=\pi/4$,
 (c)$K=1.00,\:\Omega=\pi/64$, (d)$K=5.00,\:\Omega=\pi/64$. 
The brighter tone means the higher value of SPA1. 
}\label{SPA11}
\label{SPA11-1}
\end{figure}

\begin{figure}[tb]
\includegraphics[width=5.5in,bb=0 0 772 292]{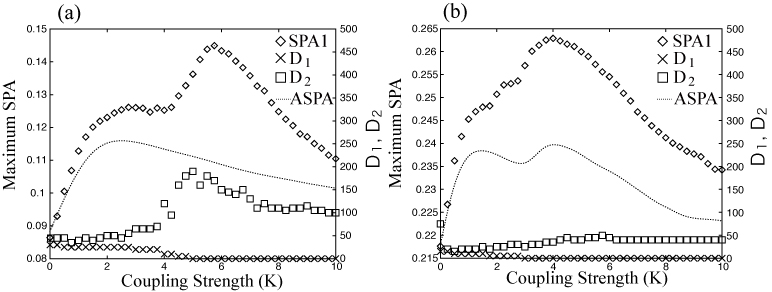}
\caption{Maximum SPA1 obtained in each K-fixed $D_1-D_2$ space
 ($\diamond$), fitting curve of Maximum ASPA (dashed line), and
 corresponding sets $(D_1,
 D_2)$ ($\Box$ and $\times$) for (a)$\Omega=\pi/4$, (b)$\Omega=\pi/64$.
 Here, only the sets satisfying $D_1 \leq D_2$ are drawn.}\label{SPA12}
\label{SPA12-1}
\end{figure}
In the next, we concentrate on the SR of particle 1.
Fig.\ref{SPA11}(a)-(d) show SPA1 (SPA of particle 1) in $D_1-D_2$ space
for different sets of $(K, \Omega)$.
The peaks in all of these figures are located {\it off} the diagonal line $D_1 = D_2$ and within the area of $D_1 < D_2$.
This means that the particle under a smaller amplitude of noise
dominantly experiences the effective SR caused by the heterogeneous
amplitude of noise (hereafter, heterogeneous SR).
Fig.\ref{SPA12}(a) and (b) show the relation between  $K$ and maximum SPA of
particle 1 (hereafter max-SPA1) in the cases of $\Omega=\pi/4$ and
$\Omega=\pi/64$ respectively,  where the sets $(D_1, D_2)$
realizing max-SPA1 in each $K$ fixed $D_1-D_2$ space also are drawn.
In the same figures, for the easiness of comparison, fitting curves of the
max-ASPA shown in Fig.\ref{ASPA2}(a) and (b) are added as dashed lines,
where the graphs of max-SPA1 start from the values same as max-ASPA at
$K=0$.
In the small $K$ region, as $K$ increases, the values of max-SPA1
monotonically increase, and after passing over shoulder parts around
which the increasing rates once decreases, they recover rapid
increasing rate until reaching maximum values.
Through whole range of $K$, max-SPA1 is kept higher than (or at least
same as) the max-ASPA of dashed lines.
This, again, means that the particles under a smaller amplitude of noise
dominantly experiences the effect of the heterogeneous SR.

If we pay attention to the $K$ value ($K \geq 5.75$) realizing the peak
of max-SPA1 in Fig.\ref{SPA12}(a) and the decreasing curve of max-ASPA
around the corresponding $K$ value, the SR of particle 1 is considered
not sufficiently significant to be reflected in the average SR of the
whole system in this case of $\Omega = \pi/4$.
Meanwhile, as shown in Fig.\ref{SPA12}(b), with the slower oscillation of
$\Omega=\pi/64$, the $K$ value giving the peak of max-SPA1 realizes the
peak of max-ASPA at the same time.
Namely, the resonance effect of particle 1 is strong enough to
overcome the poor resonance of particle 2 in this
case of slow oscillation ($\Omega = \pi/64$) of external field.
In such way, heterogeneous amplitude of noise leads the system in a
certain range of conditions, not only locally but also in the average of
the whole, to an effective SR.
\begin{figure}[b]
\includegraphics[width=4.2in, bb=0 0 702 625]{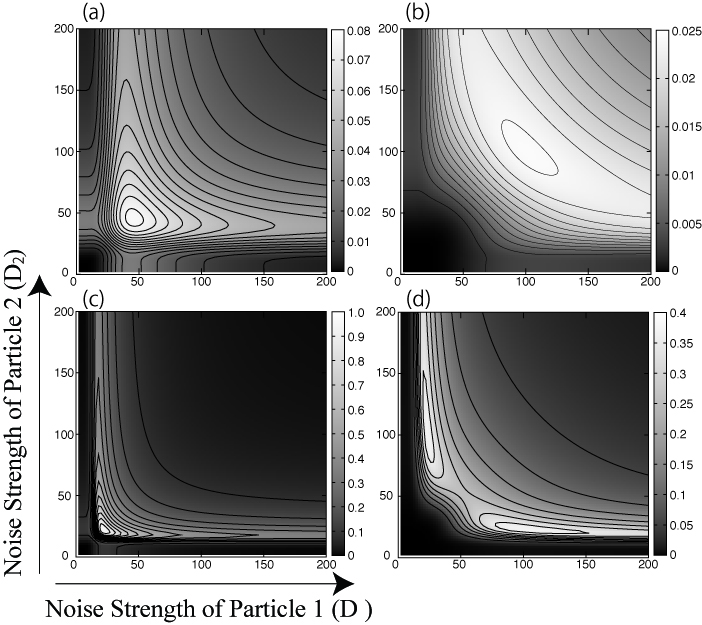}
\caption{Contour lines of ASPA in $D_1-D_2$ space for (a)$K=1.00$,
 $\Omega=\pi/4$, (b)$K=5.00$,
 $\Omega=\pi/4$, (c)$K=1.00$, $\Omega=\pi/64$, (d)$K=5.00$,
 $\Omega=\pi/64$, $A=10.0$, $\tau_0=2\pi/ \sqrt{(| -2a+k| (4a+k)}$, $\Delta=64.0$ obtained from the
 analytical calculation of master equation (7).
 The brighter tone means the higher value of ASPA.
ASPA is symmetric with respect to $D_1$ and $D_2$ from the definition
 of ASPA.
In the case of (d), the ASPA has two peaks and the peaks fall
 ${\it off}$ the diagonal line.}
\label{ising1}
\label{ising1-2}
\end{figure}

\section{Analysis Using Master Equation Model}
To understand the mechanism of the present effective SR of coupled two
particles under heterogeneous amplitude of noise, we analytically treat the
simplified transient dynamics among finite number of states using a
master equation;
\begin{eqnarray}
&&{\frac{dP(\sigma_1,\sigma_2,t)}{dt}} \\
=&&W(-\sigma_1, \sigma_2, t)P(-\sigma_1,\sigma_2,t)-W(\sigma_1,\sigma_2, t)P(\sigma_1,\sigma_2,t) \nonumber 
+W(-\sigma_2,\sigma_1, t)P(\sigma_1, -\sigma_2,t) -W(\sigma_2, \sigma_1,
  t)P(\sigma_1, \sigma_2,t),
\label{eq:master}
 \end{eqnarray}
where $\sigma_i \in \{s,-s\}$ represents the state in which
particle $i \in \{1,2\}$ is at the left minima ($\sigma_i=-s$) or the
right minima ($\sigma_i=s$) of the potential expressed by
(2) ($s=\sqrt{a/2b}$).
$W(\sigma_i, \sigma_j, t)$ denotes the transition rate
from state $(\sigma_i,  \sigma_j)$ to $(-\sigma_i, \sigma_j)$, 
the specific form of which is assumed as,
\begin{eqnarray}
W(\sigma_i, \sigma_j, t) =\frac{1}{2\tau_i}\left\{1 - \sigma_i \sigma_j
					    \tanh(K/D_i)\right\}
\left\{1-\sigma_i \tanh(A \cos \Omega t/D_i)\right\} \hspace{1pc} (i\neq
j) ,
\end{eqnarray}
where $A$, $K$, $D_i$ are the amplitude of the input signal, the
coupling strength and the noise strength added to the particle $i$, respectively.
The quantity $\tau_i$ is the Arrhenius-type relaxation time, $\tau_i=\tau_0
\exp(\Delta/D_i)$, where $\tau_0=2\pi/ \sqrt{(| -2a+k| (4a+k)}$ and $\Delta$ is the
activation energy. 
%
%
This transition rate has a similar form as that introduced in the
kinetic Ising model for $D_1=D_2$ \cite{suzuki}.
Here, we should note that this master equation model does not take into
account the direct transition between states $(\sigma_1,\sigma_2)$ and
$(-\sigma_1,-\sigma_2)$.
It indicates that this model has a direct correspondence to (5)
only at the weak coupling between two particles.

As shown in Appendix, master equation (7) is analytically
solved with several approximations introduced in ref(9) for $D_1=D_2$ though
some details of calculation are different for the present case of $D_1
\neq D_2$.
Fig.\ref{ising1}(a)-(d) show ASPA obtained using the analytical solution of
(7) in $D_1-D_2$ space with different sets of $(K, \Omega)$.

In Fig.\ref{ising1}(a) and (b), the peak of ASPA falls on the line
$D_1=D_2 > 0$ which means the equivalent noise leads the system to the most
effective SR  in $D_1-D_2$ space. 
This figures qualitatively corresponds to Fig.\ref{ASPA1}(a) and (b). 
With slower oscillation $\Omega=\pi/64$  and weak coupling $K=1.00$ 
the peak of ASPA is still on the diagonal line $D_1=D_2 > 0$ as shown in
Fig.\ref{ising1}(c), however the peaks of ASPA fall {\it off} the
diagonal line with stronger coupling  $K=5.00$ (Fig.\ref{ising1}(d))
meaning that heterogeneous amplitude of noise drives the system to the
most effective SR in  $D_1-D_2$ space.
In this way, as long as looking the contours of ASPA, 
qualitative features of Fig.\ref{ASPA1}(a)-(d) obtained by
the simulation of Langevin dynamics model (5) seems to be reproduced by the
analytical solution of master equation (7).
However some remarkable aspects of the heterogeneous SR are not sufficiently described by 
the present form of master equation model as discussed in the followings.


\begin{figure}[tb]
\includegraphics[width=5.5in,bb=0 0 762 287]{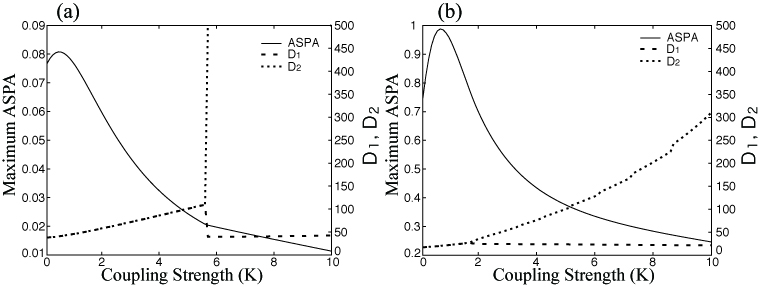}
\caption{Maximum ASPA obtained in each K-fixed $D_1-D_2$ space (solid line), and corresponding sets $(D_1, D_2)$ ($D_1$ : dashed line,
$D_2$: broken line)  for  (a)$\Omega=\pi/4$, (b)$\Omega=\pi/64$.
Here, only the sets satisfying
$D_1 \leq D_2$ are drawn and  $A=10.0$, $\tau_0=2\pi/ \sqrt{(| -2a+k| (4a+k)}$,
 $\Delta=64.0$. The curve of $D_2$  in (a) comes  out of the graph over a  critical value of $K$.}  
\label{ising2-1}
\end{figure}

To discuss the quantitative aspects, max-ASPAs obtained at each
$K$-fixed $D_1 - D_2$ space are connected in Fig.\ref{ising2-1} as graphs with varying $K$. 
Solid lines in Fig.\ref{ising2-1}(a), (b) show the relation between
$K$ and max-ASPA in the cases of $\Omega = \pi/4$ and $\Omega = \pi/64$,
respectively. 
Here, SR of the system is reinforced as the coupling between particles
is strengthened until max-ASPAs reach peak values at finite $K$
($K \simeq 0.6$ in Fig.\ref{ising2-1}(a) and $K \simeq 0.7$ in Fig.\ref{ising2-1}(b)), and thereafter
monotonically decrease. 
In the same figures, dashed and broken lines respectively give
$D_1$ and $D_2$ which collapse on each other in a range of small $K$ and
separate beyond $\Omega$-dependent critical values of $K$. 
The effect of the heterogeneous amplitude of noise
exceeds that of equivalent noise as $K$ increases.
These features shown in Fig.6 especially for $K<3$  reflect the
qualitative characters obtained by Langevin dynamics model (Fig.2) for 
the same $K$ range.
However, it should be cared in Fig.6(b), that  the effect of
heterogeneous amplitude of noise is not sharp enough to make the second
mound of the  max-ASPA graph shown in Fig.2(b) of Langevin dynamics
model, nor the overlap of $D_1$ and $D_2$ in Fig.2(b) giving max-ASPA
for large $K$ is recognized in Fig.6(b).
Moreover the dominant region of heterogeneous SR for large $K$ in
Fig.6(a) has no counterpart in Fig.2(a).


\section{Discussion and Summary}

These distinctions between behaviors of the Langevin dynamics model and
the master equation model would partially originate from the assumption of weak
coupling limit as mentioned in the previous section. 
However the more intrinsic origin for this distinction seems linked to
the basic mechanism of the present heterogeneous SR as explained below;


If we rewrite the Langevin dynamics of (5) like,
\begin{equation}
\begin{array}{l}
\displaystyle\frac{\partial x_1}{\partial t} = -\frac{\partial
 V_{\rm{eff}}(x_1,x_2)}{\partial x_1} +\xi_1(t), \\[3mm]
\displaystyle\frac{\partial x_2}{\partial t} = -\frac{\partial
V_{\rm{eff}}(x_1.x_2)}{\partial x_2}+\xi_2(t),
\label{1-2}
\end{array}
\end{equation}
 an overall effective potential  $V_{ \rm{eff} }(x_1,x_2)$ for particles 1 and 2  is, \\
 \begin{equation}
 V_{ \rm{eff} }(x_1,x_2)=a(x_1^2+ x_2^2)+b(x_1^4+x_2^4)+K(x_1-x_2)^2-A\cos(\Omega t)(x_1+x_2).
\end{equation}
If we take $x_2$ as constant in the above form,  an effective potential of particle 1  
becomes monostable above a critical coupling strength $K_c$ that depends
on $\Omega t$ and  $x_2$.
Then  the signal and the coupling force act in the same direction  
 if $\Omega t=0$,  $x_2=s(=\sqrt{a/2b}),$ (or if $\Omega t=\pi$ and
 $x_2=-s(=-\sqrt{a/2b})$), supposing coupling strength satisfies
 $K>K_c$, in which case the signal is superthreshold.
This means that particle 1 will move quickly into the single minimum (if
 it is not already there) without surmounting a potential barrier.
Noise is not necessary for this transition of particle 1.
On the other hand, particle 2 should be sufficiently noisy to trigger
 the monostable potential for particle 1.
In particular, the mean escape time of particle 2 should be much smaller
 than half the period of the signal.
In this case, particle 1 can exhibit essentially a classical (i.e.,
 deterministic) resonance, which results in a large ASPA at unequal
 noise amplitudes. In the present simulation of Langevin dynamics model,
 the critical coupling strength $K_c$ is estimated as $K_c \simeq 2.4$ using
 the specific values of parameters: $a=8.0, b=0.25, A=10.0$.

This scenario explains the split of max-ASPA graphs over  $K\simeq 2.4
\simeq K_c$ in Fig.2(b), one is max-ASPA within whole $D_1-D_2$ space and the other is that on $D_1=D_2$ line. 
Also $K_c$ is close to the initiating point of the right-side mound of max-ASPA in the same figure.
Moreover, it  should be noted that, in the limit $K \rightarrow \infty$,  max-ASPA in $D_1-D_2$ space degenerates on a line 
$D_1+D_2$= constant,  causing the vanishment of heterogeneous SR in this
limit,  which is reflected in Fig.2(b) as the rejoining of $D_1$ and
$D_2$ at large $K$ accompanied with the rejoining of max-ASPA values in
whole $D_1-D_2$ space and along  $D_1=D_2$ line in the same figure.

To summarize, we numerically showed that the
efficiency of SR is enhanced by the coupling of bistable elements 
exposed to heterogeneous amplitude of noise.
This heterogeneous SR is caused through a 'task allocation' between two
particles such that one particle (particle 2) under large fluctuation 
easily jumps over the potential barrier and pulls another particle(particle 1) to the potential minimum state keeping itself under
a large fluctuation, then only the latter (particle 1) directly  contributes to the heterogeneous SR. 
Therefore if we focus only on the SR of particle 1, it is found that the  
present mechanism works for a wide range in $K$ (coupling strength) and
$\Omega$ (frequency of external signal)  as seen in Fig.4, where the
maximum value of  SPA1, an SR index of particle 1,  is marked under
heterogeneous amplitude of noise in almost all range of $K$ in the
graph.

However, the most noticeable in the present study is that 
the effect of heterogeneous SR is largely pronounced when coupling
strength between two particles exceeds a critical value and 
the external  signal for particle 1 reaches  super-threshold level with
the help of highly fluctuating particle 2. 
In that case not only  SPA1; an index of SR of particle 1, but also
ASPA; an index of averages degree of SR of particles 1 and 2, mark the
maximum value at each-$K$ under heterogeneous amplitude of noise.  

In this way, the present heterogeneous SR is an outcome of an exquisite
combination of the triggering particle with a large fluctuation and the
following particle which faithfully responds to the external signal
under small (or zero) amplitude of noise. 
We suppose that the heterogeneous SR has potential applicability to a
wider variety of stochastic resonance phenomena like those of devising
artificial sensors with high susceptibility\cite{kawa}, and that it has a
certain connection to the basic study on the function of non-isothermal
multi-elements systems\cite{kawa}.
\acknowledgements
This study was supported by Grant-in-Aid (19654056, 22540391) for Scientific
Research,  and, the Global COE Program G14 (Formation and Development of
Mathematical Sciences Based on Modeling and Analysis) of JSPS.

\section*{Appendix}
\renewcommand{\theequation}{A.\arabic{equation}}
\setcounter{equation}{0}

In this appendix, we analytically derive the time evolution of the
expectation values of $\sigma_i$ treated in (7).
For simplification we assume $\sigma_i \in \{1, -1\}$.
From equation (7), we obtain the dynamics for the expectation values,
 \begin{eqnarray}
  \tau_1 \dot {\langle \sigma_1 \rangle}&=&  - \langle \sigma_1 \rangle +
					       \tanh \left( \beta_1 A \cos (\Omega t)\right)+ \eta_1 \langle \sigma_2
					       \rangle- \eta_1 \tanh \left( \beta_1 A \cos (\Omega t)\right)\langle
					       \sigma_1 \sigma_2 \rangle, \\
  \tau_2 \dot {\langle \sigma_2 \rangle} &=& - \langle \sigma_2 \rangle +
						\tanh \left( \beta_2 A \cos (\Omega t)\right)+ \eta_2 \langle \sigma_1
						\rangle - \eta_2 \tanh \left( \beta_2 A \cos (\Omega t)\right)\langle
						\sigma_1 \sigma_2 \rangle , \\
  \dot{\langle \sigma_1 \sigma_2 \rangle} &=&
   \frac{1}{\tau_1}\left( -\langle \sigma_1 \sigma_2 \rangle + \eta_1 +
		      \tanh(\beta_1 A \cos (\Omega t)) \langle \sigma_2
		      \rangle -\eta_1 \tanh(\beta_1 A \cos(\Omega t)) \langle
		      \sigma_1 \rangle \right) \\ \nonumber
   && + \frac{1}{\tau_2}\left( -\langle \sigma_1 \sigma_2 \rangle + \eta_2 +
			   \tanh(\beta_2 A \cos (\Omega t)) \langle \sigma_1
			   \rangle -\eta_2 \tanh(\beta_2 A \cos (\Omega t)) \langle
			   \sigma_2 \rangle \right),
 \end{eqnarray}
where $\eta_i \equiv \tanh \left( K \beta_i \right)$ and $\beta_i \equiv
1/D_i$.
To solve the equation, we assume that input signal is sufficiently
weak ($\beta_i A \ll 1$), then, the correlation function $\langle
\sigma_1 (t) \sigma_2 (t) \rangle$ has even parity with respect to $A$. 
Consequently, only the zeroth order term in $A$ remains in this
correlation in the linear approximation. 
An additional assumption is that $\langle \sigma_1 (t) \sigma_2 (t)
\rangle$ is in a steady state, which means that $\langle \sigma_1 (t)
\sigma_2 (t) \rangle$ is expressed like,
\begin{equation}
 \langle \sigma_1 \sigma_2 \rangle = \frac{\eta_1 \tau_2 + \eta_2
  \tau_1}{\tau_1 + \tau_2}.
\end{equation}

Then (A1) and (A2) are simplified like, 
\begin{eqnarray}
\displaystyle \left[
\begin{array}{ccc}
\dot {\langle \sigma_1 \rangle} \\
\dot {\langle \sigma_2 \rangle} \\
\end{array}
\right]
&=&
\left[
\begin{array}{ccc}
-\frac{1}{\tau_1} \hspace{0.5pc} \frac{\eta_1}{\tau_1} \\
\frac{\eta_2}{\tau_2} \hspace{0.5pc}-\frac{1}{\tau_2} \\
\end{array}
\right]
\left[
\begin{array}{ccc}
\langle \sigma_1 \rangle\\
\langle \sigma_2 \rangle\\
\end{array}
\right]
+\left[
\begin{array}{ccc}
H_1 \cos (\Omega t)\\
H_2 \cos (\Omega t)\\
\end{array}
\right],\\ \nonumber
\displaystyle H_1 &=& \frac{\left( 1-\eta_1 \langle \sigma_1 \sigma_2 \rangle
		      \right) \beta_1 A}{\tau_1}, \\ \nonumber
\displaystyle H_2 &=& \frac{\left( 1-\eta_2 \langle \sigma_1 \sigma_2 \rangle
		      \right) \beta_2 A}{\tau_2}. \\ \nonumber
\end{eqnarray}


The long time limit solution of the above equation is
\begin{eqnarray}
 \langle \sigma_i (t) \rangle_{\rm{longtime}} &=& R_{i1} \cos(\Omega t -
  \varphi_1) -
  R_{i2} \cos (\Omega t - \varphi_2) = R_i \cos (\Omega t - \psi_i)
  \hspace{1pc} ( i \in \{ 1, 2\}),
\end{eqnarray}
where, 
\begin{eqnarray}
\lambda_1 &=& \frac{1}{2} \left( -\left(\rho_1 + \rho_2 \right)+
				 \sqrt{\left( \rho_1 - \rho_2 \right)^2 +
				 4\eta_1 \eta_2 \rho_1 \rho_2 }
			  \right),  \nonumber \\
\lambda_2 &=& \frac{1}{2} \left( -\left(\rho_1 + \rho_2 \right)-
				 \sqrt{\left( \rho_1 - \rho_2 \right)^2 +
				 4\eta_1 \eta_2 \rho_1 \rho_2 }
			  \right), \nonumber \\
L &=& 2 \eta_1 \rho_1, \nonumber \\
M &=& \left( \rho_1 - \rho_2 \right)+  \sqrt{\left( \rho_1 - \rho_2 \right)^2 +
				 4\eta_1 \eta_2 \rho_1 \rho_2 },  \nonumber \\
N &=& \left(\rho_1 - \rho_2 \right)-
				 \sqrt{\left( \rho_1 - \rho_2 \right)^2 +
				 4\eta_1 \eta_2 \rho_1 \rho_2 },  \nonumber \\
R_{11} &=& \frac{NH_1 - LH_2}{(N-M) \sqrt{\lambda_1^2 + \Omega^2}},  \nonumber\\
R_{12} &=& \frac{-MH_1 + LH_2}{(N-M) \sqrt{\lambda_2^2 + \Omega^2}},
 \nonumber\\
R_{21} &=& \frac{M(NH_1 - LH_2)}{L(N-M) \sqrt{\lambda_1^2 + \Omega^2}},
 \nonumber \\
R_{22} &=& \frac{N(-MH_1 + LH_2)}{L(N-M) \sqrt{\lambda_2^2 +
\Omega^2}}, \nonumber \\
R_i &=& \sqrt{ R_{i1}^2 + R_{i2}^2 + 2R_{i1}R_{i2} \cos (\varphi_1
 -\varphi_2)}, \nonumber \\
\varphi_i &=& \tan^{-1} \left( \frac{\Omega}{\lambda_i}\right),  \nonumber \\
 \psi_i &=& \tan^{-1} \left( \frac{R_{i1}\sin\varphi_1 + R_{i2}
		       \sin \varphi_2}{R_{i1}\cos\varphi_1 + R_{i2}
		       \cos \varphi_2} \right), \nonumber \\
\rho_i &\equiv& \frac{1}{\tau_i},  \nonumber \\
\Gamma_i &\equiv& \exp(\lambda_i t - \lambda_i t_0).  \nonumber 
\end{eqnarray}

Hence, the auto-correlation function is
\begin{eqnarray}
\langle \sigma_i(t+\Delta t) \sigma_i(t)\rangle_{\rm{longtime}} =
R_{i}^2 \cos(\Omega t + \Omega \Delta t- \psi_i) \cos(\Omega t -
\psi_i),
\end{eqnarray}
and averaging it over initial phase, we get
\begin{eqnarray}
\langle \langle \sigma_i(t+\Delta t) \sigma_i(t)\rangle \rangle &=&
 \frac{\Omega}{2 \pi} \int^{2 \pi /\Omega}_{0} \langle \sigma_i(t+\Delta
 t) \sigma_i (t)\rangle_{\rm{longtime}} dt \nonumber \\
&=& \frac{R_{i}^2 \cos(\Omega \Delta t)}{2}.
\end{eqnarray}

To obtain the power spectra, we apply the Wiener-Khinchin theorem to
(A.8). The power spectrum $S_i (\omega)$ defined by
\begin{equation}
 S_i (\omega) = \int^{\infty}_{-\infty} e^{-i \omega \Delta t} \langle
  \langle \sigma_i(t+\Delta t) \sigma_i(t) \rangle \rangle d \Delta t ,
\end{equation}
is calculated in the form
\begin{equation}
S_i (\omega) = \frac{\pi R_{i}^2}{2} \Bigl[\delta (\Omega - \omega) + \delta
 (\Omega + \omega) \Bigr].
\end{equation}

And the power spectrum about input signal ($A\cos(\Omega t)$) is written as 
\begin{eqnarray}
   S_{in} (\omega) &=& \int^{\infty}_{-\infty} \langle A\cos(\Omega t + \Omega \Delta
  t) A\cos (\Omega t) \rangle \exp(-i\omega \Delta t) d \Delta t
  \nonumber \\
&=& \frac{\pi A^2}{2} \left(\delta (\Omega - \omega) + \delta (\Omega +
		       \omega)\right).
\end{eqnarray}

By substituting $S_i$ for $S_{out}$ in (4), we obtain the local SPA,
i.e. SPA$_i$ $(i \in \{1, 2\})$, and the average SPA, i.e. ASPA, 
\begin{eqnarray}
 SPA_i &=& \frac{R_i^2}{A^2}, \\
ASPA &=& \frac{SPA_1 + SPA_2}{2} = \frac{R_1^2 + R_2^2}{2 A^2}.
\end{eqnarray}

In the present case of $\sigma \in \{ s, -s\}$, to comparing with
Langevin Equation (5) $R_i (K, A)$ is written as $R_i(s^2K, sA)$, the
expectation value $\sigma_i$ and SPA also are simply rewritten as
\begin{eqnarray}
 \langle \sigma_i \rangle_{\rm{longtime}} &=& s R_i(s^2K, sA) \cos(\Omega t -
  \psi_i), \\
SPA_i &=& \frac{s^2 R_i(s^2K, sA)^2}{A^2}.
\end{eqnarray}

\end{document}